\documentclass{emulateapj}

\usepackage[backref,breaklinks,colorlinks,citecolor=blue]{hyperref}
\usepackage[all]{hypcap}
\usepackage{apjfonts}

\usepackage{url}
\slugcomment{Accepted for publication in ApJ}

\shorttitle{Lower Bound on the TeV Gamma-ray Background}
\shortauthors{Inoue \& Tanaka}

\begin{document}
\title{Lower Bound on the Cosmic TeV Gamma-ray Background Radiation} 

\author{Yoshiyuki Inoue\altaffilmark{1}} 
\affil{Institute of Space and Astronautical Science JAXA, 3-1-1 Yoshinodai, Chuo-ku, Sagamihara, Kanagawa 252-5210, Japan}
\email{yinoue@astro.isas.jaxa.jp}

\author{Yasuyuki T. Tanaka} 
\affil{Hiroshima Astrophysical Science Center, Hiroshima University, 1-3-1 Kagamiyama, Higashi-Hiroshima, Hiroshima 739-8526, Japan}

\altaffiltext{1}{JAXA International Top Young Fellow}

\begin{abstract}
The {\it Fermi} gamma-ray space telescope has revolutionized our understanding of the cosmic gamma-ray background radiation in the GeV band. However, investigation on the cosmic TeV gamma-ray background radiation still remains sparse. Here, we report the lower bound on the cosmic TeV gamma-ray background  spectrum placed by the cumulative flux of individual detected extragalactic TeV sources including blazars, radio galaxies, and starburst galaxies. The current limit on the cosmic TeV gamma-ray background above 0.1 TeV is obtained as $2.8\times10^{-8} (E/100~{\rm GeV})^{-0.55} \exp(-E/2100~{\rm GeV})~{\rm [GeV/cm^2/s/sr]} < E^2dN/dE < 1.1\times10^{-7} (E/100~{\rm GeV})^{-0.49}~{\rm [GeV/cm^2/s/sr]}$, where the upper bound is set by requirement that the cascade flux from the cosmic TeV gamma-ray background radiation can not exceed the measured cosmic GeV gamma-ray background spectrum \citep{ino12}. Two nearby blazars, Mrk~421 and Mrk~501, explain $\sim70$\% of the cumulative background flux at 0.8--4~TeV, while extreme blazars start to dominate at higher energies. We also provide the cumulative background flux from each population, i.e. blazars, radio galaxies, and starburst galaxies which will be the minimum requirement for their contribution to the cosmic TeV  gamma-ray background radiation.
\end{abstract}

\keywords{gamma rays: diffuse background - gamma rays: general - cosmic background radiation}

\section{Introduction}
\label{intro}
The {\it Fermi} gamma-ray space telescope (hereinafter {\it Fermi}) has successfully measured the cosmic gamma-ray background (CGB) spectrum at 0.1-820 GeV \citep{ack15_cgb}. The CGB represents superposed gamma-ray flux from all resolved and unresolved gamma-ray sources in the universe outside of the Milky way\footnote{The cosmic gamma-ray background (CGB) is also called as the extragalactic gamma-ray background (EGRB or EGB) or the isotropic gamma-ray background (IGRB) where the EGB is the total (resolved and unresolved) CGB and the IGRB is the unresolved CGB \citep[e.g.][]{ack15_cgb}. In other wavelengths, it is common to use the term of the cosmic background radiation such as the cosmic microwave, infrared, optical, and X-ray background radiation. }.  In this paper, we simply refer to the total CGB as the CGB otherwise noticed. 

Various gamma-ray emitting sources have been discussed as the origins of the CGB in the literature \citep[see][for recent reviews]{ino14_fermi,for15}. {\it Fermi} enables us to understand its composition at $>$0.1~GeV \citep{aje15,mau15} as blazars \citep[e.g.][]{aje12,aje14}, radio galaxies \citep[e.g.][]{ino11}, and star-forming galaxies \citep[e.g.][]{ack12_sfg}. At $>50$~GeV, source count analysis based on the {\it Fermi} source catalog detected above 50 GeV \citep[2FHL;][]{ack15_2fhl} found that current detected populations make up $86^{+16}_{-14}$~\% of the CGB and its source counts are compatible with the expected blazar source counts \citep{dim15}.

The cosmic TeV gamma-ray background, however, has not been well investigated yet. At the TeV energy region, ground based gamma-ray telescopes such as imaging atmospheric Cherenkov telescopes (IACTs) observe gamma rays through the air shower produced by the gamma ray interacting with the atmosphere. Since hadrons and electrons also produce air shower, those background events need to be subtracted. In the standard analysis procedure, the background flux level is determined using regions of no gamma-ray emitting objects but in the same field of view \citep{ber07}. This method subtracts the CGB emission signals together with other hadronic and leptonic backgrounds. It is therefore difficult to measure the isotropic diffuse CGB radiation with this method, although the Galactic diffuse emission has recently been measured by the H.E.S.S. collaboration \citep{abr14_diffuse}.

As an aside, the IceCube Collaboration has recently reported detection of several tens of TeV--PeV neutrino events \citep{aar13,aar14}. The origin of the IceCube neutrinos are still under debate \citep[see e.g.][for reviews]{mur15}. Conventionally, those high energy neutrinos are produced by cosmic rays via hadronuclear ($pp$) and/or photohadronic ($p\gamma$) interactions. In either case, gamma rays are accompanied with. The current unresolved CGB spectrum at the GeV band constrains $pp$ scenarios as the origin of IceCube TeV--PeV neutrino events \citep{mur13,bec15} because a power-law secondary spectrum following the initial cosmic-ray spectrum is generated. As gamma-ray and neutrino spectra of $p\gamma$ scenarios depend on target photon densities \citep[e.g.][]{mur14,der14}, spectral extrapolation like the $pp$ models is not valid. Therefore, the cosmic TeV--PeV gamma-ray background spectrum would be useful to constrain neutrino origins further. However, gamma rays above $\sim100$~GeV propagating through the universe experience absorption by the interaction with the extragalactic background light (EBL) via electron--positron pair production \citep[e.g.][]{gou66,jel66,ste92,fin10, dom11, ino13_cib}. This EBL attenuation may severely suppress TeV gamma-ray signals from neutrino origins, while neutrinos are not suppressed.

In this paper, we place the lower bound on to the cosmic TeV gamma-ray background spectrum. Current IACTs have detected 131 sources \citep[TeVcat;][]{tevcat}\footnote{http://tevcat.uchicago.edu/} of which $\sim50$ sources are extragalactic objects, blazars, radio galaxies, and starburst galaxies. Integration of low-state flux of those extragalactic TeV sources provides a firm lower limit on the cosmic TeV gamma-ray background radiation. This method is an analogy of galaxy counts which integrate the flux of individual detected galaxies and gives a lower bound on to the EBL \citep[e.g.][]{mad00,tot01}. We also show the cumulative flux of each population and the allowed range of the cosmic TeV gamma-ray background radiation together with the upper bound which is placed not to make the GeV cascade component of the cosmic TeV gamma-ray background radiation exceed the measured unresolved CGB spectrum \citep{ino12}.

\section{Extragalactic TeV Source Samples}

\capstartfalse
\begin{deluxetable*}{lccccccr}
\tabletypesize{\scriptsize}
 \tablecaption{Extragalactic TeV Gamma-ray Objects at $|b|\ge10$~deg\label{tab:sample}}
\tablehead{
  \colhead{Source} & \colhead{R.A. [deg]}  & \colhead{Dec. [deg]} & \colhead{l [deg]} & \colhead{b [deg]} & \colhead{Class} & \colhead{3FGL~Name}  & \colhead{Reference}\
 }
\startdata
SHBL~J001355.9-185406 & 3.48 & -18.90 & 74.63 & -78.08 &  blazar    & 3FGL~J0013.9-1853 &   \citet{hess13_SHBLJ001355.9-185406}\\
NGC~253 & 11.89 & -25.27 & 97.50 & -87.95 & starburst galaxy & 3FGL~J0047.5-2516 &   \citet{abr12_ngc253}\\
RGB~J0152+017 & 28.16 & 1.81 & 152.36 & -57.52 &  blazar    & 3FGL~J0152.6+0148 &   \citet{aha08_RGBJ0152+017}\\
3C~66A & 35.67 & 43.03 & 140.15 & -16.77 &  blazar   & 3FGL~J0222.6+4301 &   \citet{ale11_3C66A}\\
1ES~0229+200 & 38.22 & 20.27 & 152.97 & -36.62 &  blazar   & 3FGL~J0232.8+2016 &   \citet{aha07_1ES0229+200}\\
 PKS~0301-243 & 45.87 & -24.12 & 214.63 & -60.17 &  blazar    & 3FGL~J0303.4-2407 &   \citet{hess13_PKS0301-243}\\
NGC~1275 & 49.96 & 41.51 & 150.58 & -13.26 & radio galaxy & 3FGL~J0319.8+4130 &   \citet{ale14_ngc1275}\\
RBS~0413 & 49.97 & 18.79 & 165.09 & -31.67 &  blazar   & 3FGL~J0319.8+1847 &   \citet{ali12_RBS0413}\\
1ES~0347-121 & 57.31 & -11.98 & 201.89 & -45.74 &  blazar   & 3FGL~J0349.2-1158 &   \citet{aha07_1ES0347-121}\\
1ES~0414+009 & 64.22 & 1.08 & 191.83 & -33.16 &  blazar   & 3FGL~J0416.8+0104 &   \citet{abr12_1ES0414+009}\\
 PKS~0548-322 & 87.66 & -32.28 & 237.58 & -26.15 &  blazar    & 3FGL~J0550.6-3217 &   \citet{hess10_PKS0548-322}\\
RGB~J0710+591 & 107.59 & 59.15 & 157.39 & 25.41 &  blazar   & 3FGL~J0710.3+5908 &   \citet{acc10_RGBJ0710+591}\\
 S5~0716+71 & 110.49 & 71.35 & 143.98 & 28.02 &  blazar   & 3FGL~J0721.9+7120 &   \citet{and08_S50716+714}\\
1ES~0806+524 & 122.45 & 52.31 & 166.26 & 32.91 &  blazar   & 3FGL~J0809.8+5218 &   \citet{acc09_1ES0806+524}\\
M~82 & 148.87 & 69.67 & 141.44 & 40.54 & starburst galaxy & 3FGL~J0955.4+6940 &   \citet{acc09_m82}\\
1RXS~J101015.9-311909 & 152.57 & -31.34 & 266.93 & 20.04 &  blazar   & 3FGL~J1010.2-3120 &   \citet{abr12_1RXSJ101015.9-311909}\\
1ES~1011+496 & 153.77 & 49.43 & 165.53 & 52.72 &  blazar   & 3FGL~J1015.0+4925 &   \citet{alb07_1ES1011+496}\\
1ES~1101-232 & 165.89 & -23.49 & 273.18 & 33.08 &  blazar   & 3FGL~J1103.5-2329 &   \citet{aha07_1ES1101-232}\\
Mrk~421 & 166.12 & 38.21 & 179.83 & 65.03 &  blazar   & 3FGL~J1104.4+3812 &   \citet{alb07_mrk421} \\
1ES~1215+303 & 184.46 & 30.12 & 188.87 & 82.05 &  blazar   & 3FGL~J1217.8+3007 &   \citet{ale12_1ES1215+303}\\
1ES~1218+304 & 185.34 & 30.18 & 186.36 & 82.73 &  blazar   & 3FGL~J1221.3+3010 &   \citet{acc09_1ES1218+304}\\
M~87 & 187.73 & 12.41 & 283.84 & 74.51 & radio galaxy & 3FGL~J1230.9+1224 &   \citet{ale12_M87}\\
1ES~1312-423 & 198.69 & -42.63 & 307.50 & 20.04 &  blazar    & 3FGL~J1314.7-4237 &   \citet{hess_1ES1312-423}\\
Cen~A & 201.37 & -43.03 & 309.52 & 19.41 & radio galaxy & 3FGL~J1325.4-4301 &   \citet{aha09_cena}\\
PKS~1424+240 & 216.76 & 23.80 & 29.49 & 68.20 & blazar  & 3FGL~J1427.0+2347 &   \citet{acc10_pks1424+240}\\
H~1426+428 & 217.15 & 42.67 & 77.47 & 64.89 &  blazar   & 3FGL~J1428.5+4240 &   \citet{aha02_H1426+428}\\
AP~Librae & 229.42 & -24.38 & 340.68 & 27.58 &  blazar   & 3FGL~J1517.6-2422 &   \citet{abr15_APLibrae}\\
PG~1553+113 & 238.94 & 11.19 & 21.92 & 43.96 & blazar  & 3FGL~J1555.7+1111 &   \citet{ale12_pg1553+113}\\
Mrk~501 & 253.48 & 39.75 & 63.59 & 38.85 &  blazar   & 3FGL~J1653.9+3945 &   \citet{alb07_Mrk501}\\
1ES~1959+650 & 300.02 & 65.15 & 98.01 & 17.67 &  blazar    & 3FGL~J2000.0+6509 &   \citet{magic06_1ES1959+650}\\
PKS~2005-489 & 302.35 & -48.83 & 350.38 & -32.60 &  blazar   & 3FGL~J2009.3-4849 &   \citet{hess10_PKS2005-489}\\
PKS~2155-304 & 329.72 & -30.23 & 17.73 & -52.25 &  blazar   & 3FGL~J2158.8-3013 &   \citet{hess10_PKS2155-304}\\
BL~Lac & 330.69 & 42.28 & 92.60 & -10.44 &  blazar   & 3FGL~J2202.7+4217 &   \citet{alb07_BLLac}\\
B3~2247+381 & 342.53 & 38.42 & 98.26 & -18.57 &  blazar   & 3FGL~J2250.1+3825 &   \citet{ale12_B32247+381}\\
H~2356-309 & 359.83 & -30.65 & 12.71 & -78.07 &  blazar   & 3FGL~J2359.3-3038 &   \citet{aha06_H2356-309}
\enddata
\end{deluxetable*}
\capstarttrue

\begin{figure}
 \begin{center}
  \includegraphics[width=8cm]{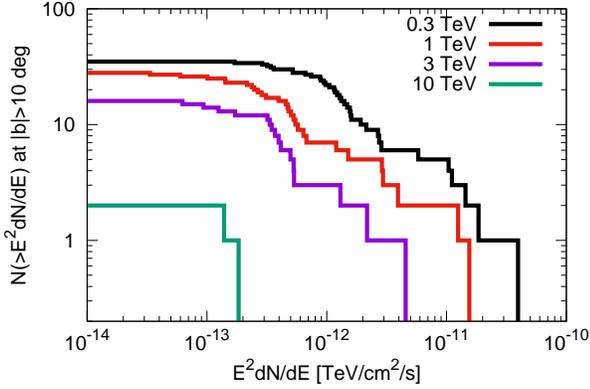} 
 \end{center}
\caption{Cumulative source count distribution of extragalactic TeV detected objects at $|b|>10$~deg as a function of energy flux at an energy indicated in the figure. The sample includes blazars, radio galaxies, and starburst galaxies. Each line corresponds to the distribution at 0.3~TeV, 1~TeV, 3~TeV, and 10~TeV from top to  bottom.}\label{fig:counts}
\end{figure}

We select 35 known extragalactic TeV sources which are located at Galactic latitude $|b|> 10$~deg and whose low activity state flux is available, since our aim is to give conservative constraints on the CGB in the TeV band. It is not straightforward to define the low-state flux with IACTs because IACTs do not always monitor sources like {\it Fermi}. Therefore, for each source, we select the lowest fluxes among several TeV measurements by modern IACTs (H.E.S.S., MAGIC, and VERITAS) and further restrict samples showing no significant variability in the TeV band during observations. The sample contains 30 blazars, 3 radio galaxies, and 2 starburst galaxies from the default TeVcat catalog \citep{tevcat} which include published sources only. Energy bins of TeV gamma-ray spectra in the literature are different among papers. To make energy bins even, we rebin TeV spectra by interpolating between each binned data in the range of reported energies. We also include the {\it Fermi} third source (3FGL) catalog data \citep{ace15_3fgl} to cover GeV gamma-ray spectra. The 3FGL catalog is based on its first 48 months of survey data. All of our sample have counterparts in the 3FGL catalog. Our sample is summarized in \autoref{tab:sample}.

\autoref{fig:counts} shows the cumulative source count distribution of our TeV source sample at 0.3~TeV, 1~TeV, 3~TeV, and 10~TeV.  As energy increases, the number of the sample decreases. The apparent distribution at each energy is different from a uniform distribution in the Euclidean universe. However, the sky coverage of current  IACTs is not uniform. Further discussion on the cumulative source count distribution at the TeV band requires more uniform and wide sky coverage by future experiments  \citep{ino10_cta} such as the Cherenkov Telescope Array \citep[CTA;][]{act11} and the High Altitude Water Cherenkov observatory \citep[HAWC;][]{abe13}.

\section{Lower Bounds on the Cosmic Gamma-ray Background Radiation}
\label{sec:bound}

\begin{figure}
 \begin{center}
  \includegraphics[width=8.5cm]{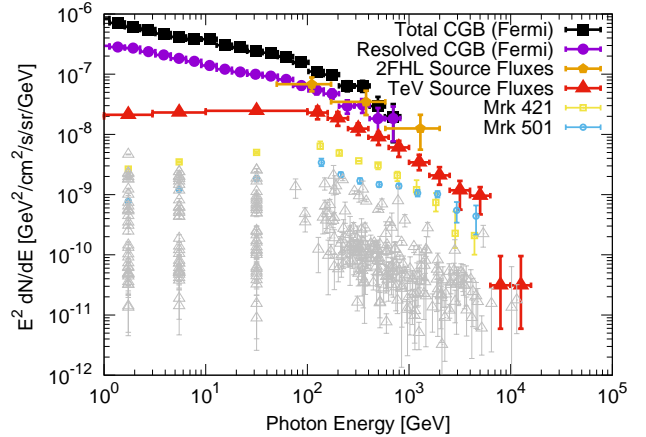} 
 \end{center}
\caption{The cosmic gamma-ray background spectrum at the GeV--TeV band. The lower bound on the cosmic TeV gamma-ray background obtained from the cumulative flux of 35 known extragalactic TeV objects at $|b|>10$~deg is shown by filled triangles. Filled square, filled circle, and filled pentagon data points represent the total CGB spectrum measured by {\it Fermi} \citep{ack15_cgb}, the resolved CGB spectrum by {\it Fermi} \citep{ack15_cgb}, and the cumulative 2FHL extragalactic source fluxes based on \citet{ack15_2fhl}. The contributions of individual objects are shown by open triangle data, where the flux is divided by the sky area at $|b|>10$~deg. Open square and circle points represent that of Mrk~421 and Mrk~501, respectively. The error bars correspond to 1-$\sigma$ uncertainty.}\label{fig:cgb_ll}
\end{figure}

The lower bound on the cosmic TeV gamma-ray background is obtained by integrating flux of our TeV samples at each energy bin. To convert this integrated  flux to the cumulative background flux, we divide the flux by the sky area above $|b|\ge 10$~deg ($\simeq3.3\pi$~str). The obtained background spectrum at each energy is tabulated in \autoref{tab:cgb}. The uncertainties at each energy is also estimated by integrating the rebinned 1-sigma upper and lower bounds of each source and by dividing those values by the sky area above $|b|\ge 10$~deg. \autoref{fig:cgb_ll} shows the lower bound on the cosmic TeV gamma-ray background radiation spectrum together with the {\it Fermi} CGB spectrum \citep{ack15_cgb}. The TeV CGB flux resolved by current IACTs is dominated by two nearby bright blazars, Mrk~421 and Mrk~501. These two objects make $\sim70$\% of the flux at 0.8--4~TeV. At $>4$~TeV,  on the other hand, extreme blazars, H1426-428 and 1ES~0229+200, both of which are detected up to $\sim10$~TeV start to dominate the background flux. 

The obtained lower bound above 100 GeV is well approximated as 
\begin{eqnarray}
\nonumber
E^2\frac{dN}{dE}&\ge&2.8_{-0.63}^{+0.72}\times10^{-8} \left(\frac{E}{100~{\rm GeV}}\right)^{-(0.55_{-0.020}^{+0.047})}\\
&&\times \exp\left(-\frac{E}{2.1_{-0.47}^{+0.80}\times10^3~{\rm GeV}}\right)~{\rm [GeV/cm^2/s/sr]},
\end{eqnarray} 
The exponential cutoff may indicate the feature of gamma-ray attenuation by EBL in the CGB spectrum. However, it can be also interpreted as the intrinsic gamma-ray spectral cutoff in individual gamma-ray sources, since Mrk~421 and Mrk~501 dominate the lower bound and the gamma-ray opacity at the distance to these two blazars becomes unity at $\sim7$~TeV \citep[e.g.][]{ino13_cib}.

 For the comparison, we also show in \autoref{fig:cgb_ll} the {\it Fermi} resolved CGB spectrum which corresponds to the cumulative flux of the {\it Fermi} detected sources in \citet{ack15_cgb}. The cumulative flux of the 2FHL catalog sources is also shown. For the 2FHL sources, we collect sources at $|b|\ge10$~deg listed in the 2FHL catalog \citep{ack15_2fhl} where 257 objects are included.   

Current IACTs have resolved $\sim30$\% of the CGB flux measured by {\it Fermi} at $\sim1$~TeV, while {\it Fermi} itself has remarkably resolved $\sim90$\% of that. The resolved fraction is consistent between \citet{ack15_cgb} and the 2FHL catalog \citep{ack15_2fhl}. Although the resolved CGB flux by IACTs is consistent with the {\it Fermi} resolved CGB flux considering uncertainties, this difference should be addressed because current IACTs have about a factor of 40 better sensitivity at 1~TeV than {\it Fermi}  \citep[see also {\url{http://www.slac.stanford.edu/exp/glast/groups/canda/lat_Performance.htm}} for the latest {\it Fermi} sensitivity][]{fun13}. 

\begin{table}
\begin{center}
\caption{The lower bound on the cosmic TeV gamma-ray background spectrum\label{tab:cgb}}
  \begin{tabular}{lc}
      \tableline
      Energy & Lower Bound Spectrum\\ 
      (GeV) & $E^2dN/dE$ (${\rm GeV/cm^2/s/sr}$) \\ 
     \tableline
      0.20 & ($1.9{\pm0.2})\times10^{-8}$ \\
      0.65 & ($2.0{\pm0.1})\times10^{-8}$ \\
      2.0 & ($2.1{\pm0.1})\times10^{-8}$ \\
      6.5 & ($2.3{\pm0.1})\times10^{-8}$ \\
      55 & ($2.5{\pm0.2})\times10^{-8}$ \\
      130 & ($2.3_{-0.5}^{+0.6})\times10^{-8}$ \\
      205 & ($1.9\pm{0.5})\times10^{-8}$ \\
      325 & ($1.2{\pm0.3})\times10^{-8}$ \\
      515 & ($9.0_{-2.3}^{+2.0})\times10^{-9}$ \\
      815 & ($6.1{\pm1.8})\times10^{-9}$ \\
      1300 & ($3.4_{-0.9}^{+1.2})\times10^{-9}$ \\
      2050 & ($2.1_{-0.6}^{+0.8})\times10^{-9}$ \\
      3250 & ($1.2_{-0.6}^{+0.5})\times10^{-9}$ \\
      5150 & ($9.6_{-4.9}^{+3.7})\times10^{-10}$ \\
      8200 & ($3.1_{-2.5}^{+6.4})\times10^{-11}$ \\
      13000 & ($3.1_{-2.5}^{+6.4})\times10^{-11}$ \\
      \tableline
    \end{tabular}
\end{center}
\end{table}

The difference of resolved fractions may be due to the sky coverage difference. {\it Fermi} covers the whole sky, while current IACTs cover only limited sky regions. In the 2FHL catalog, $\sim78$\% of the sources which are detected at $>50$~GeV are not observed by IACTs. However,  only 14 objects are listed at the highest energy bin (585--2000~GeV) at $|b|\ge10$~deg in the 2FHL catalog and our sample includes 30 sources at energies above 585~GeV. Therefore, the difference of the sky coverages might not be the main cause of this resolved fraction difference. This implies a few bright objects dominate the cosmic TeV gamma-ray background flux as the resolved component is dominated by Mrk~421 and Mrk~501. 

Another interpretation is temporal variability. The CGB spectrum is the time-averaged spectrum of all over the sky taken by the first 50~months operation of the {\it Fermi}. The 2FHL catalog averages the source variation in 80 months of data. As blazars are highly variable \citep[e.g.][]{abd10_var}, average flux is expected to be higher than low-state flux. For example, the gamma-ray flux of Mrk~421 at 585--2000~GeV is $5.1_{-1.3}^{+1.6} \times10^{-11}~{\rm ph/cm^2/s}$ and $1.9_{-0.37}^{+0.50}\times10^{-11}~{\rm ph/cm^2/s}$ in the 2FHL catalog and our sample data \citep{alb07_mrk421}, respectively. Since we have collected low-state data only to put a conservative lower bound on the the cosmic TeV gamma-ray background radiation, the obtained bound can be lower than the averaged 2FHL and {\it Fermi} resolved CGB flux. This flux difference of Mrk~421 alone would reconcile the difference between the 2FHL source flux and the IACTs source flux. 

The gamma-ray flux of the other dominant blazar Mrk~501 at 585--2000~GeV is $1.1_{-0.51}^{+0.75}\times10^{-11}~{\rm ph/cm^2/s}$ and $1.5_{-0.16}^{+0.17}\times10^{-11}~{\rm ph/cm^2/s}$ in the 2FHL catalog and our sample data \citep{alb07_Mrk501}, respectively, which are relatively similar to each other. It is known that Mrk~421 shows more frequent variabilities than Mrk~501 \citep{abd11_mrk421,abd11_mrk501}. Continuous monitoring and understanding of duty cycle of extragalactic gamma-ray sources especially Mrk~421 are necessary to obtain the averaged gamma-ray flux and the resolved CGB flux.

\begin{figure}
 \begin{center}
  \includegraphics[width=8.5cm]{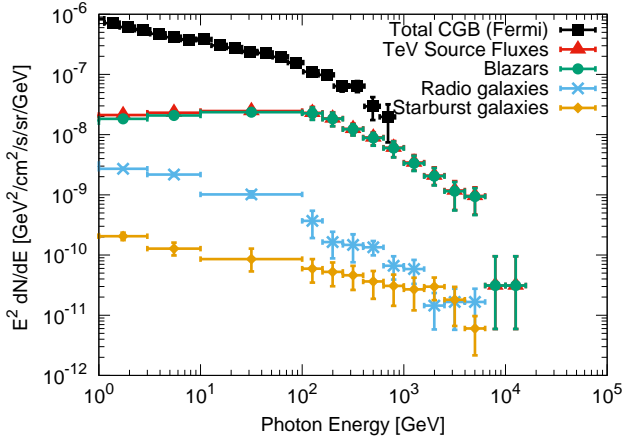} 
 \end{center}
\caption{Cumulative gamma-ray background spectrum from various populations detected in the TeV band. Circle, cross, and diamond points correspond to cumulative flux of detected blazars, radio galaxies, and starburst galaxies at $|b|>10$~deg, respectively.  The lower bound of the cosmic TeV gamma-ray background from the cumulative flux of those populations is shown by triangle. Square data points represent the total CGB spectrum measured by {\it Fermi} \citep{ack15_cgb}. The error bars correspond to 1-$\sigma$ uncertainty. As the TeV source counts are dominated by blazars, circle and triangle points almost overlap each other.}\label{fig:cgb_pop}
\end{figure}

\autoref{fig:cgb_pop} shows the cumulative TeV gamma-ray spectrum from each population; blazars, radio galaxies, and starburst galaxies. As 30 of 35 samples are blazars, they dominate the cumulative flux of currently known extragalactic TeV sources. In the GeV band, it is well-known that blazars dominate the cosmic gamma-ray background radiation \citep[e.g.][]{ino09,aje12,aje14,aje15,mau15}, while radio galaxies \citep[e.g.][]{ino11,dim14} and starburst galaxies \citep[e.g.][]{pav02, fie10, mak11,ack12_sfg, lac14} make sub-dominant contributions. The contribution of radio galaxies and starburst galaxies is approximated by a power-law spectral shape with an index of -0.9 and -0.5 in $E^2dN/dE$, respectively.

\section{Discussion and Conclusion}

In this paper, we obtain the lower bound on the cosmic TeV gamma-ray background spectrum. The bound is set by the cumulative flux of known TeV gamma-ray objects at $|b|\ge10$~deg. We collect low-state flux data of 35 TeV gamma-ray emitting objects. By including the 3FGL data catalog, the bounds is set from 100~MeV to $\sim10$~TeV. The bound is dominated by two nearby blazars Mrk~421 and Mrk~501 which make $\sim70$\% of the resolved CGB by IACTs at $\sim$0.8--4~TeV. However, at higher energies, extreme blazars which are more distant than Mrk~421 and Mrk~501 start to dominate. Comparing with the CGB spectrum measured by {\it Fermi}, current known TeV sources explain $\sim30$\% of the {\it Fermi} CGB flux at $\sim1$~TeV.

\begin{figure}
 \begin{center}
  \includegraphics[width=8cm]{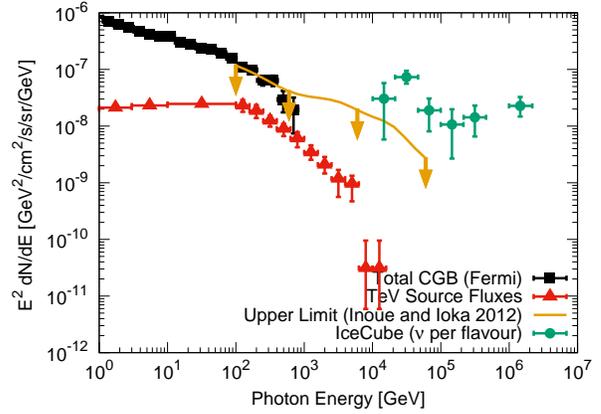} 
 \end{center}
\caption{The cosmic background photon and neutrino spectra from GeV to PeV. The square, triangle, and circle data points represent the total CGB spectrum measured by {\it Fermi} \citep{ack15_cgb}, the cumulative flux of known extragalactic TeV objects at $|b|>10$~deg (the lower bound), and the IceCube neutrino flux per flavour \citep{aar14}, respectively. The solid curve with arrows represents the upper bound on the CGB requiring cascade emission not to exceed the EGB data below 100 GeV in the model-independent way \citep{ino12}. Error bars correspond to 1-$\sigma$ uncertainty of data.}\label{fig:cgb_limit}
\end{figure}

Here, EBL photons attenuate gamma rays through the electron--positron pair production process. Those pairs scatter the cosmic microwave background (CMB) radiation via the inverse Compton scattering  and generate secondary gamma-ray emission component \citep[the so-called cascade emission; e.g.][]{pro86,aha94,fan04}.\footnote{Although plasma beam instability may suppress the cascade emission \citep{bro12}, recent Particle-In-Cell simulations reveals that the plasma instability carries 10\% of the attenuated energy at most \citep{sir14}.} The cascade component must contribute to the cosmic GeV gamma-ray background radiation, if there is the cosmic TeV gamma-ray background radiation  \citep{cop97,mur12_cgb, ino12, ack15_cgb}. The level of the upper bound weakly depends on the assumed spectral energy distribution and evolution of contributors to be consistent with the {\it Fermi} measurements \citep{ino12}. By setting no evolution, a power-law emission with a photon index of 1.5, and a cutoff energy of 60~TeV, the current unresolved CGB measurement below 100 GeV sets an upper bound on the CGB itself at $>100$~GeV as $1.1\times10^{-7} (E/100~{\rm GeV})^{-0.49}~{\rm GeV/cm^2/s/sr}$ \citep{ino12}. 

Combining with the lower and upper bounds, the allowed range of the cosmic TeV gamma-ray background spectrum is approximated as $2.8\times10^{-8} (E/100~{\rm GeV})^{-0.55} \exp(-E/2100~{\rm GeV})~{\rm [GeV/cm^2/s/sr]} < E^2dN/dE < 1.1\times10^{-7} (E/100~{\rm GeV})^{-0.49}~{\rm [GeV/cm^2/s/sr]}$. \autoref{fig:cgb_limit} shows the current bounds on the cosmic TeV gamma-ray background radiation together with the IceCube neutrino flux per flavour  \citep{aar14}. As the current {\it Fermi} unresolved CGB measurements give constraints on the origin of the TeV-PeV neutrino background (e.g. \citet[][]{mur13,bec15}, but see also \citet{mur15_pp,kis15}), our bounds at the TeV band may be useful for the further constraints. However, it should be noted that gamma-ray attenuation by the EBL photons suppresses the associated gamma-ray signals \citep[e.g.][]{fin10,dom11,ino13_cib,kha15}, while neutrinos are not suppressed. Moreover, if neutrinos and gamma rays are generated in dense environments like starforming galaxies, TeV gamma rays can be internally attenuated by pair production because of luminous interstellar radiation photon field \citep[e.g.][]{dom05,ino11_sb,mur15,kis15}.

Next generation ground gamma-ray telescopes CTA will have a factor of $\sim10$ better sensitivity than current IACTs \citep{act11}. And, HAWC which covers over 5~str of the sky will achieve better sensitivity and wider energy coverage than current IACTs do \citep{abe13}. Once these observatories perform extragalactic surveys, more TeV sources are expected to be detected \citep{ino10_cta,dub13}. Future CTA and HAWC sky surveys will tighten the current lower bound further. If the cosmic TeV gamma-ray background flux is close to our lower bound and dominated by a few objects as our results show, strong anisotropy signature can be expected.

At $\gtrsim10$~TeV, the lower bound on the cosmic TeV gamma-ray background seems to be flat in $E^2dN/dE$, although the flux uncertainty is still large. At these energy band, extreme blazars make up the cumulative flux rather than nearby bright blazars. Extreme blazars do not show apparent variabilities and hard gamma-ray spectra and the one-zone synchrotron self-Compton model fits requires extreme parameters  \citep[see ][and references therein]{tan14}. 
 Various models have been proposed to explain extreme blazars such as the stochastic acceleration scenarios \citep{lef11} and the lepto-hadronic emission scenario \citep{cer15}. Exotic scenarios are also discussed such as hypothetical axion-like particles \citep{dea07,sim08,san09}, as well as  Lorentz invariance violation \citep{kif99,pro00}. 
 
 An alternative interpretation for extreme blazars is the cascade emission from high energy cosmic rays propagating through intergalactic space. Protons or neutrons escaping from the jet initiate cascades with EBL or CMB \citep[e.g.][]{ess10,mur12,tak13,ess14}. In the cascade scenario, the observed fluxes contain two emission components: primary gamma-ray flux produced at the source and secondary gamma-ray flux, which arises from line-of-sight interactions of cosmic rays during the propagation. A secondary gamma-ray component from the cascade scenario creates a flat spectrum above several TeV to a few tens of TeV. Despite EBL attenuation, the cascade scenarios allow us to detect many blazars even at $>1$~TeV beyond the EBL attenuation horizon \citep{ino14_cas}. Flat signature in the cosmic TeV gamma-ray background radiation will be a key for the test of the cosmic-ray induced cascade scenario and for the understanding of neutrino origins as this cascade process also produces TeV-PeV neutrinos \citep{ess11,kal13}.

\bigskip
The authors thank the anonymous referee for useful comments and suggestions. The authors would also like to thank Marco Ajello, John Beacom, and Mattia Di Mauro for useful comments and discussions. Y.I. acknowledges support by the JAXA international top young fellowship. YTT is supported by Kakenhi 15K17652.

\end{document}